\documentclass[fleqn,usenatbib]{mnras}
\usepackage{graphicx}
\usepackage{natbib}
\usepackage{epstopdf}
\usepackage{amssymb}
\usepackage{array, makecell}
\usepackage{amsmath}
\usepackage{mathptmx}
\usepackage{booktabs}
\usepackage{hyperref}
\usepackage{float}
\usepackage[T1]{fontenc}

\DeclareRobustCommand{\VAN}[3]{#2}
\let\VANthebibliography\thebibliography
\def\thebibliography{\DeclareRobustCommand{\VAN}[3]{##3}\VANthebibliography}
\usepackage{graphicx}	
\usepackage{amsmath}	
\usepackage{amssymb}	

\usepackage[normalem]{ulem}
\usepackage{xcolor}
\definecolor{new_color}{rgb}{1.0, 0.43, 0.3}

\newcommand {\tg}{\ensuremath{\,r_{\rm g} / c}}

\newcommand{\uu}[1]{_{_{\rm #1}}}

\title[GRMHD simulations of BH activation by small scale loops]{GRMHD simulations of BH activation by small scale magnetic loops: Formation of
striped jets and active coronae}

\author[A. Chashkina, O. Bromberg, A. Levinson]{Anna Chashkina,$^{1}$\thanks{E-mail: chashkina.anna@gmail.com}
Omer Bromberg,$^{1}$ and Amir Levinson$^{1}$
\\
$^{1}$The Raymond and Beverly Sackler School of Physics and Astronomy, Tel Aviv University, Tel Aviv 69978, Israel\\
}

\date{Accepted XXX. Received YYY; in original form ZZZ}

\pubyear{2021}

\begin{document}
\label{firstpage}
\pagerange{\pageref{firstpage}--\pageref{lastpage}}
\maketitle

\begin{abstract}
We have performed a series of numerical experiments aimed at studying the activation of Kerr black holes (BHs) by
advection of small scale magnetic fields. Such configurations may potentially give rise to the formation of
quasi-striped Blandford-Znajek jets. It can also lead to enhanced dissipation and generation of plasmoids in
current sheets formed in the vicinity of the BH horizon, which may constitute a mechanism to power the
hard X-ray emission seen in many accreting BH systems (a la lamppost models).  Our analysis suggests
that formation of quasi-striped jets with significant power may be possible provided loops 
with alternating polarity 
having sizes larger than $\sim 10 r_g$ or so 
can be maintained (either form sporadically or advected from
outside) at a radius $\lesssim 10^2 r_g$.  This conclusion is consistent with recent results of general relativistic force-free simulations.  
We also find that the accretion dynamics exhibits cyclic behaviour in MAD states, alternating between high accretion phases and 
quenched accretion phases during which the magnetosphere becomes force-free out to radii $\gtrsim 10r_g$. We suggest that such a behaviour should lead to notable variations of the observed luminosity and image of the inner disc (BH shadow image).
Finally, we find that the transition between accreted loops on the BH gives rise to the formation of current sheets and energetic plasmoids on the jet boundary during intermittent periods when the jet becomes inactive, in addition to an equatorial current sheet that forms during peaks in the jet activity.
\end{abstract}




\section{Introduction}

Much progress has been made in recent years in the study of relativistic jets.   The exquisite EHT data \citep{EHT2019p1,EHT2021} in combination
with a large bank of sophisticated GRMHD simulations \citep{komissarov09,Tchekhovskoy11,Tchekhovskoy12,Bromberg2016,sadowski16,Porth2019,vourellis19,Liska2020}
has lent support to the hypothesis that relativistic jets are launched by 
magnetic extraction of the rotational energy of a Kerr black hole (for a recent review see \citealt{blandford19} and references therein).   However, the question as to how the magnetic energy is 
ultimately being converted to the hot, dense plasma emitting  the observed radiation is far from being resolved.  
Magnetic field dissipation ultimately requires formation of current sheets and turbulence inside the jet.  There are two ways to achieve this;
the first one requires generation of global instabilities that can lead to a strong distortion of magnetic field lines.  The second one
involves advection of small scale magnetic fields by the accretion flow that can, potentially, lead to the launching of a quasi-striped jet.  

It has been established that advection of large scale magnetic fields by the hot accretion flow can lead to the efficient production of a powerful jet
(particularly in the Magnetically Arrested Disc (MAD) regime), but the stability of such jets has not been widely explored (particularly under realistic conditions).
The current driven kink instability has been identified as a potential mechanism to generate strong distortions that can ultimately lead to
formation of current sheets and turbulence, however, under which conditions this instability develops, and at what scales, is yet unclear. 
Numerical experiments devoted to the study of relativistic kink instability in Poynting flux
dominated jets \citep{Mizuno2012,Bromberg2019,Davelaar2019}, 
seem to indicate that strong collimation, or even focusing, is essential for a rapid growth of the instability \citep{Bromberg2016}.
Such conditions are favourably anticipated in GRBs inside the star \citep{Aloy00,Bromberg2016,obergaulinger17}, 
or in reconfinement 
zones occasionally seen in AGNs (e.g., the HST-1 knot in M87).  However, these zones are typically located far from the black hole (BH)
at radii $10^5-10^7\ r_g$ from the putative  BH, where $r_g$ is the gravitational radius,  while in many objects dissipation is seen or inferred on much smaller scales.
One should keep in mind, though, that these numerical experiments are still not sufficiently advanced to be applied to realistic setups.

The issue of magnetic field dissipation close to the BH is also of utmost relevance to the hard X-ray emission observed in some 
Seyfert galaxies and stellar-mass BH systems (many of which do not exhibit jets).  This emission appears to originate from 
a compact corona located at a few
gravitational radii above the BH, consistent with the so-called "lamppost" geometry (e.g., \citealt{uttley14,kara16,wilkins11,ursini20}).  
However, the nature of this corona is yet unknown.
A plausible power source likely involves reconnection of small scale magnetic fields in the vicinity of the horizon \citep{Yuan19}.  This process 
can ultimately lead to inverse Compton emission by either hot (or non-thermal) electrons continuously heated (or accelerated) 
by the dissipation process or, alternatively, bulk motions of multi-scale plasmoids generated in the reconnecting current sheets \citep{sironi20}. 
This provides another incentive to study the structure and dynamics of the inner magnetosphere in such setups.

Advection of small scale magnetic fields into the BH  can naturally lead to substantial dissipation of magnetic energy on horizon scales. The question is whether such a configuration also allows formation of a relativistic jet with substantial power. 
This question has been addressed recently using 2D \citep{Parfrey15} and 3D \citep{Yuan19,mahlmann20} 
general relativistic force-free simulations (GRFFE), that considered
advection of magnetic loops with alternating polarity.    These numerical models demonstrate formation of intermittent,  quasi-striped jets aligned with the rotation of the BH
, as well as formation of current sheets and plasmoids near the jet boundary.   While the mean efficiency of the jet 
formation process is considerably smaller than in the case of ordered magnetic fields (MAD in particular), it is nonetheless substantial, 
ranging from about 0.1 to 0.4 of  the stationary energy extraction by the Blandford-Znajek process (BZ). 
The disc in those simulations is treated as a boundary condition in the equatorial plane that advects magnetic 
loops into the ergosphere, and it remains unclear whether the relatively high jet power seen in most runs is imposed by the
artificial conditions near the equatorial plane.    
In this paper we present a complimentary approach by studying
the effect of small scale magnetic fields accretion with GRMHD simulations. We run both 2D and 3D simulations and study the efficiency of jet launching and the dissipation processes that take place in the vicinity of the BH.
   
We stress that none of these approaches can capture the essential physics of this process, since 
the force-free simulations invoke an artificial disc as a boundary condition, while in GRMHD simulations the
initial setup, particularly the initial magnetic structure, is not self-consistently generated
and the boundary condition at the outer disc surface is not well constrained.    Nevertheless, comparing these two methods can be instructive.

\section{Numerical model}
The 2D and 3D simulations presented in this paper were performed using the GRMHD code {\tt HARMPI} \citep{harmpi} - a modified version of the {\tt HARM} code \citep{Gammie2003,Noble2006}. The simulations are carried out in spherical Boyer-Lindquist coordinates $r, \theta, \phi$. The initial condition in all the numerical experiments described below include a standard \citet{Fishbone1976} torus around a Kerr black hole with a Kerr parameter $a$. The inner edge of the torus is located at $r_{\rm in}=15r_g$, and the pressure maximum at $r_{\rm max}=32r_g$. The simulation box extends out to a radius $r_{\rm out}=10^5r_g$ for the 3D simulations, which is larger than the light travel distance at the duration of our 3D runs.  Thus, we expect the influence of the boundaries on the results to be insignificant.  In the 2D case the simulation domain is smaller, with an outer
radius $r_{\rm out}=120 r_g$.  Nonetheless, as will be shown, despite the relatively small box size the boundary effects are small.

 \begin{figure}
        \centering
        \includegraphics[width=0.5\textwidth]{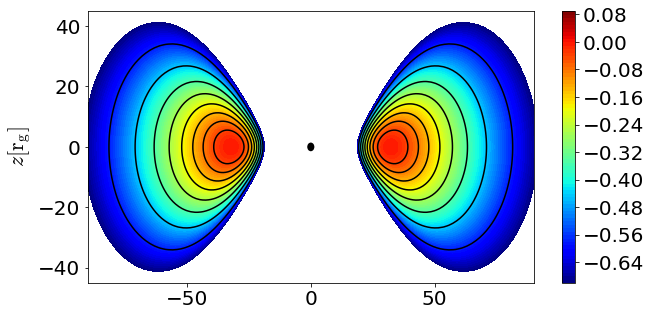}
        \includegraphics[width=0.5\textwidth]{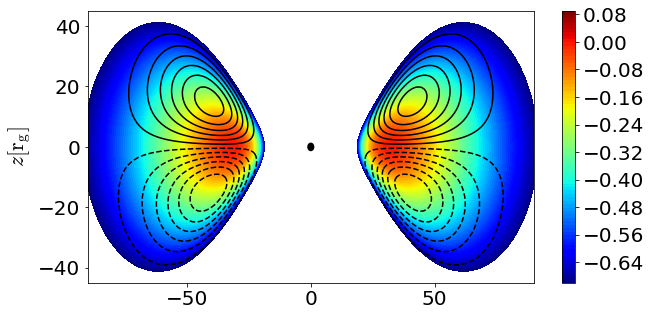}
        \includegraphics[width=0.5\textwidth]{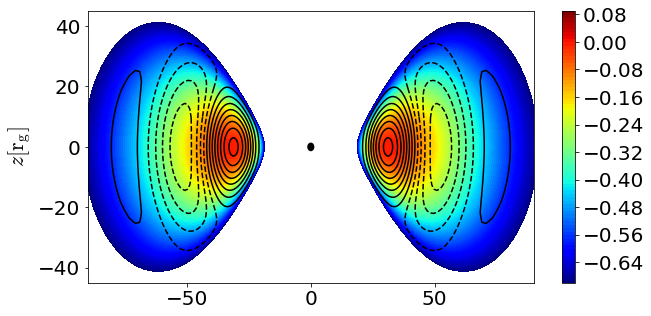}
        \includegraphics[width=0.5\textwidth]{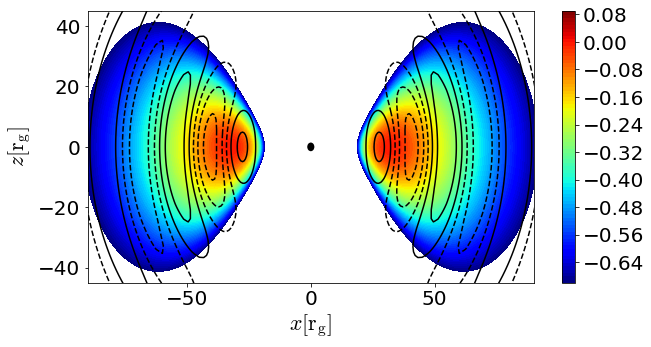}
        \caption{The initial magnetic field configurations used in our simulations, from top to bottom: one dipole, quadruple, three dipole and multiple loops cases. Magnetic field lines (represented by equipotential surfaces of the vector potential $A_{\phi}$) are shown with black lines. Solid and dashed lines indicate different polarity. Colors show the density ($\log \rho)$, see color bar.  }
        \label{fig:figure1}
    \end{figure}

The basic resolution is $N_r\times N_{\theta} \times N_{\phi} = 256\times 128\times 64$ for the 3D models and $N_r\times N_{\theta} \times N_{\phi} = 512\times 512 \times 1$ for the 2D models.  We distributed the grid in the radial direction as follow: for $r<r_{\rm br}=100\, {\rm r_g}$ we use a logarithmically spaced grid with ${\rm d}r/r={\rm constant}$, and for $r>r_{br}$ the grid scales as ${\rm d}r/r=4 (\log r)^{3/4}$. 
The initial magnetic field in the torus is specified in terms of a vector potential $A_{\phi}$. We consider three different initial magnetic field configurations, shown in Fig.~\ref{fig:figure1}:

\begin{enumerate}
    \item A single loop:
    \begin{equation}
        A_{\phi}=A_0(\rho-\rho_{\rm{cut}}),
    \end{equation}
    here $\rho_{\rm cut}=0.2$ is a threshold density below which the magnetic field vanishes.
    \item Multiple quadruple loops:
    \begin{equation}\label{eq:2}
        A_{\phi}=A_0(\rho-\rho_{\rm{cut}})\sin(\pi r_{\rm norm}N_{\rm loops}) \cos(\theta),
    \end{equation}
    where $N_{\rm loops}$ is the number of the loops, and 
    \begin{equation}
      \displaystyle  r_{\rm norm}=\displaystyle\frac{\displaystyle r/r^{\rm torus}_{\rm in}-1}{r^{\rm torus}_{\rm out}/r^{\rm torus}_{\rm in}-1},
    \end{equation}
with $r^{\rm torus}_{\rm in}$, $r^{\rm torus}_{\rm out}$ denoting, respectively, the inner and outer radii of the torus (or more precisely its magnetized section, defined by $\rho > \rho_{\rm cut}$).   
\item Multiple dipole loops:
we use a similar approach but without $\cos(\theta)$:
\begin{equation}
        A_{\phi}=A_0(\rho-\rho_{\rm{cut}})\sin(\pi r_{\rm norm}N_{\rm loops}).
    \end{equation}
\end{enumerate}

The normalization parameter $A_0$ is chosen to ensure the required initial ratio of gas to magnetic pressure $\beta=p_g/p_m\geqslant  \beta_{min}$. 
 
The various 2D models are listed in Table~\ref{tab:2D_simulations} and the 3D models in Table~\ref{tab:3D_simulations}. 
We adopt a BH spin of $a=0.9$, with the exception of model {\tt 2D1d} for which $a=-0.9$ (counter rotating disc), and an equation of state of an ideal relativistic gas (adiabatic index $\gamma=4/3$), except in model {\tt 2D1a} ($\gamma=5/3$).

\begin{table*}
    \centering
    \begin{tabular}{c|c|c|c|c|c|c}
        Model Name & Magnetic field configuration & $a/M$ & $\beta_{\rm min}$ & $\Psi_{\rm init}$ & $\Delta t_{\rm av}$, ${\rm r_{g}/c}$ & comments  \\
        \hline
        {\tt \href{https://www.youtube.com/watch?v=rUefVEAzfPs}{2D1}} & 1 loop & $0.9$ &  $1$ &  $18.08$ & [2000;7800] & fiducial 2D model \\
         \hline
         {\tt \href{https://youtu.be/aEovRzU_i6s}{2D1a}} & 1 loop & $0.9$ &  $1$ &  $21.65$ & [1000;9800] & $\gamma=5/3$  \\
         \hline
         {\tt \href{https://youtu.be/DFK-FPR6X3g}{2D1b}} & 1 loop & $0.9$ &  $100$ & $0.95$ & [3400;10000] & \\
         
         \hline
         {\tt \href{https://youtu.be/by4MLr76Euw}{2D1c}} & 1 loop & $0.9$ & $1$ & $25.57$ & [1200;6700] & $r_{\rm out}=10^4{\rm r_{g}}$\\ 
         \hline
         {\tt \href{https://youtu.be/aIVEtozArHs}{2D1d}} & 1 loop & $-0.9$ & $1$ & $11.43$ & [1200;8000]\\ 
         \hline
         {\tt \href{https://www.youtube.com/watch?v=eQcvWEnM2G0}{2D2}}  & 2 quadruple loops & $0.9$ & $1$ & $11.27$ & [1800;10000] & \\
         \hline
         {\tt \href{https://www.youtube.com/watch?v=Bk-OgPn-lQc}{2D3}} & 3 dipole loops & $0.9$ & $1$ & $14.40$ & [500;9000] & \\ 
         \hline
          {\tt \href{https://youtu.be/tn_-DWlpdlE}{2D3H}} & 3 dipole loops & $0.9$ & $1$ & $14.35$ & [600;9500] & the resolution is $2048 \times 2048$\\ 
         \hline
         
    \end{tabular}
    \caption{Details of the 2D simulations used in this paper. The initial configurations are shown in Fig.~\ref{fig:figure1}. The averaging time  $\Delta t_{\rm av}$ is defined as the time interval during which the disc mass ranges between $99-20\%$ of its initial mass. Links to movies of the simulations are available in the electronic version, click on the model name to reach the associated movie.}
    \label{tab:2D_simulations}
\end{table*}

\begin{table*}
    \centering
    \begin{tabular}{c|c|c|c|c|c|c}
        Model Name & Magnetic field configuration & $a/M$ & $\beta_{\rm min}$ & $\Psi_{\rm init}$ & $\Delta t_{\rm av}$, ${\rm r_{g}/c}$ & simulation time \\
        \hline
        {\tt \href{https://www.youtube.com/watch?v=c_49677KZQ8}{3D1}} & 1 loop & $0.9$ & $1$ & $25.94$ & [1800;10000] & $10000$ \\
        \hline
         {\tt \href{https://www.youtube.com/watch?v=uax-DHSIoZ0}{3D2}} & 2 quadruple loops & $0.9$ & $1$ & $18.81$ & [600;9000] & $22350$ \\
        \hline
         {\tt \href{https://www.youtube.com/watch?v=Vxw3PLhebYk}{3D3}} & 3 dipole loops & $0.9$ &  $1$ & $23.98$ & [800;10000] & $10000$ \\
         \hline
         {\tt \href{https://www.youtube.com/watch?v=0d0AdnhlCu8}{3DM}} & Multiple loops & $0.9$ & $1$ & $5.79$ & [1100;5200] & $11000$ \\
         \hline
    \end{tabular}
    \caption{Details of the 3D simulations used in this paper. The initial configurations are shown in Fig.~\ref{fig:figure1}. The averaging time  $\Delta t_{\rm av}$ is defined as the time interval during which the disc mass ranges between $99-20\%$ of its initial mass. Links to movies of the simulations are available in the electronic version, click on the model name to reach the associated movie.}
    \label{tab:3D_simulations}
\end{table*}

\section{Results}
\subsection{2D fiducial model}

\begin{figure}
    \centering
    \includegraphics[width=0.5\textwidth]{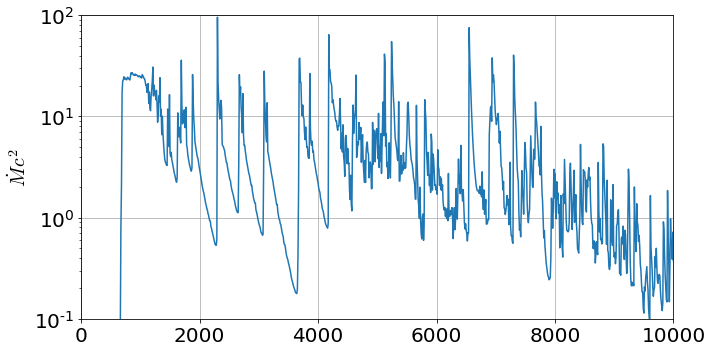}
     \includegraphics[width=0.5\textwidth]{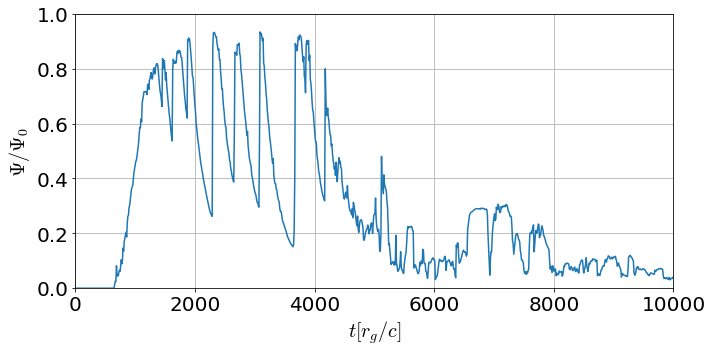}
    \caption{The evolution of the mass accretion rate (upper panel) and  dimensionless magnetic flux (lower panel) through the horizon in model {\tt 2D1}. At  early stages, magnetic flux is accumulated and reaches saturation approximately at  $t = 1500 \, \tg$. The accretion flow becomes magnetically arrested and is correlated with the flux. Accretion flow switches from MAD to SANE at $t\sim 4500\,\tg$.}
    \label{fig:figure2}
\end{figure}

We begin by analyzing the fiducial 2D one-loop simulation (model {\tt 2D1} in Table \ref{tab:2D_simulations}). The results of all other simulations will be compared with this one. The torus is initially dominated by gas pressure with $\beta \geqslant \beta_{min}=1$\footnote{Note that despite the fact that $\beta_{min}$ is small, the average $\beta$ in the simulation is of the order of few tens}. At the beginning of the simulation the magnetic field is amplified by an axisymmetric magneto-rotational instability (MRI) mode \citep{MRI}, which leads to a redistribution of angular momentum in the disc allowing for accretion to begin. We note that in both our 2D and 3D simulations MRI is well resolver in the equatorial plane, as the number of cells per MRI wavelength both in $r$ and $\theta$ directions is high enough ($Q_{r,\theta} \sim 100$).

The time evolution of the system is shown in Fig. \ref{fig:figure2}. The upper panel shows the evolution of the mass accretion rate through the BH horizon $\dot M=-\iint\limits_{\theta,\phi}\rho u^r {\rm d}A_{\Omega}$, where $\rho$ is the proper mass density, $u^r$ is the $r$-component of contravariant 4-velocity, ${\rm d}A_{\Omega}=\sqrt{-g}{\rm d}\theta {\rm d}\phi$ is a solid angle differential area at a given $r$ and $g$ is the metric determinant. 
The middle panel shows the evolution of magnetic flux through the horizon, $\Psi=\displaystyle \frac{1}{2} \iint\limits_{\theta, \phi}B^r {\rm d}A_{\Omega}$, normalized by the initial magnetic flux $\Psi_0$ stored in the disc. 
The magnetic flux on the horizon saturates shortly after the onset of the accretion at $t\simeq 1000\,\tg$. It maintains a high (fluctuating) value with an average  of $\Psi/\Psi_0=0.6$ during the first $t\simeq 4500 \tg$ then drops sharply to an average value of  $\Psi/\Psi_0=0.12$ throughout the rest of the simulation. We identify the first stage with MAD (Magnetically Arrested Disc) and the second with 
SANE (Standard And Normal Evolution).

\begin{figure*}
    \centering
    \includegraphics[width=\textwidth]{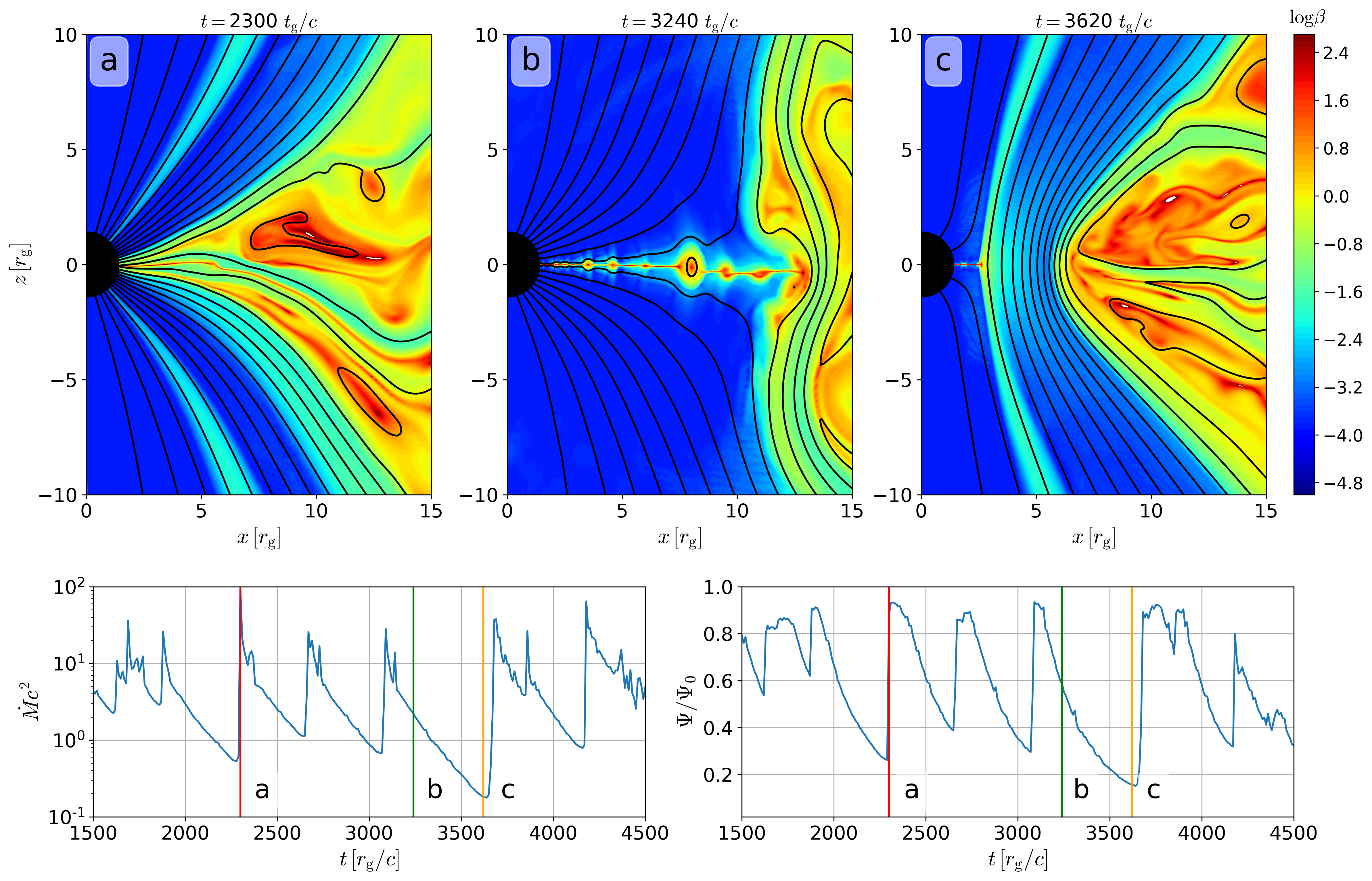}
    \caption{The accretion dynamics in a MAD state in model {\tt 2D1}. The three upper panels show snapshots of $\log \beta$. Magnetic field lines are shown in black lines. The lower panels show the mass accretion rate at the horizon (left) and normalized magnetic flux through horizon (right). Vertical lines indicate the times of the snapshots.
    }
    \label{fig:figure3}
\end{figure*}

A MAD state in 2D simulations is characterized by a saw-tooth behavior of the mass accretion rate and of the magnetic flux. In Fig.~\ref{fig:figure3} we show three snapshots of the density profiles in $r-\theta$ plane, corresponding to different parts of the saw-tooth pattern. 
The minima of the mass accretion rate are pure MAD states when accretion is halted due  by magnetic strains. 
During the dips in mass accretion rate the magnetic flux on the BH horizon also reaches a minimum, due to reconnection on the equatorial plane that detaches field lines leading to the "floating magnetic barrier" seen in Fig. \ref{fig:figure3}c. Ram pressure buildup on the inner edge of the arrested disc eventually overcomes the magnetic barrier and pushes the detached field lines back to the horizon, leading to a jump in the magnetic flux and a sharp increase in the accretion rate (Fig. \ref{fig:figure3} a).
The rapid accretion is made possible along magnetic field lines directly connecting the accretion disc with the BH horizon. Once the flux spikes, magnetic pressure pushes back and starts another phase of reconnection on the equator which reduces the flux on the horizon and the accretion rate until it is halted again and the cycle restarts (Fig. \ref{fig:figure3} b).
The decrease in horizon magnetic flux and accretion rate is exponential with a typical decay time scale of $\sim 140 \tg$, which is steady throughout the simulation. We identify this time with the reconnection time scale ($r/v_{_{\rm rec}}$), where the reconnection speed is known to be of the order of a few per cent of the Alfven speed  $v_{_{\rm rec}}\sim 0.01v_{\rm A}$ \citep{Philippov2020}.
The reconnection episodes are accompanied by plasmoid production, as proceeds during the whole exponential decay stage. 

The cyclic behaviour of the accretion dynamics suggests that the emission characteristics should vary over a timescale of tens to hundreds
dynamical times ($r_g/c$). In M87 this translates to a duration of weeks to months. 
Phase (a) in Fig.  \ref{fig:figure3} exhibits regular accretion, during which low frequency emission (radio-to-optical) is expected from the very inner regions
($r\lesssim 10 r_g$).  In contrast, in phase (b) the inner magnetosphere ($r < 10r_g$) is completely force-free, with an equatorial current sheet.
In this phase we anticipate a lower luminosity and a vastly different radio image. Dissipation in the current sheet, on the other hand, might lead to enhanced hard X-ray emission. 
Our 3D simulations, described below, show a similar behaviour albeit with a smaller contrast between the phases.  

\subsubsection{Loop sizes}

\begin{figure*}
    \centering
    \includegraphics[width=\textwidth]{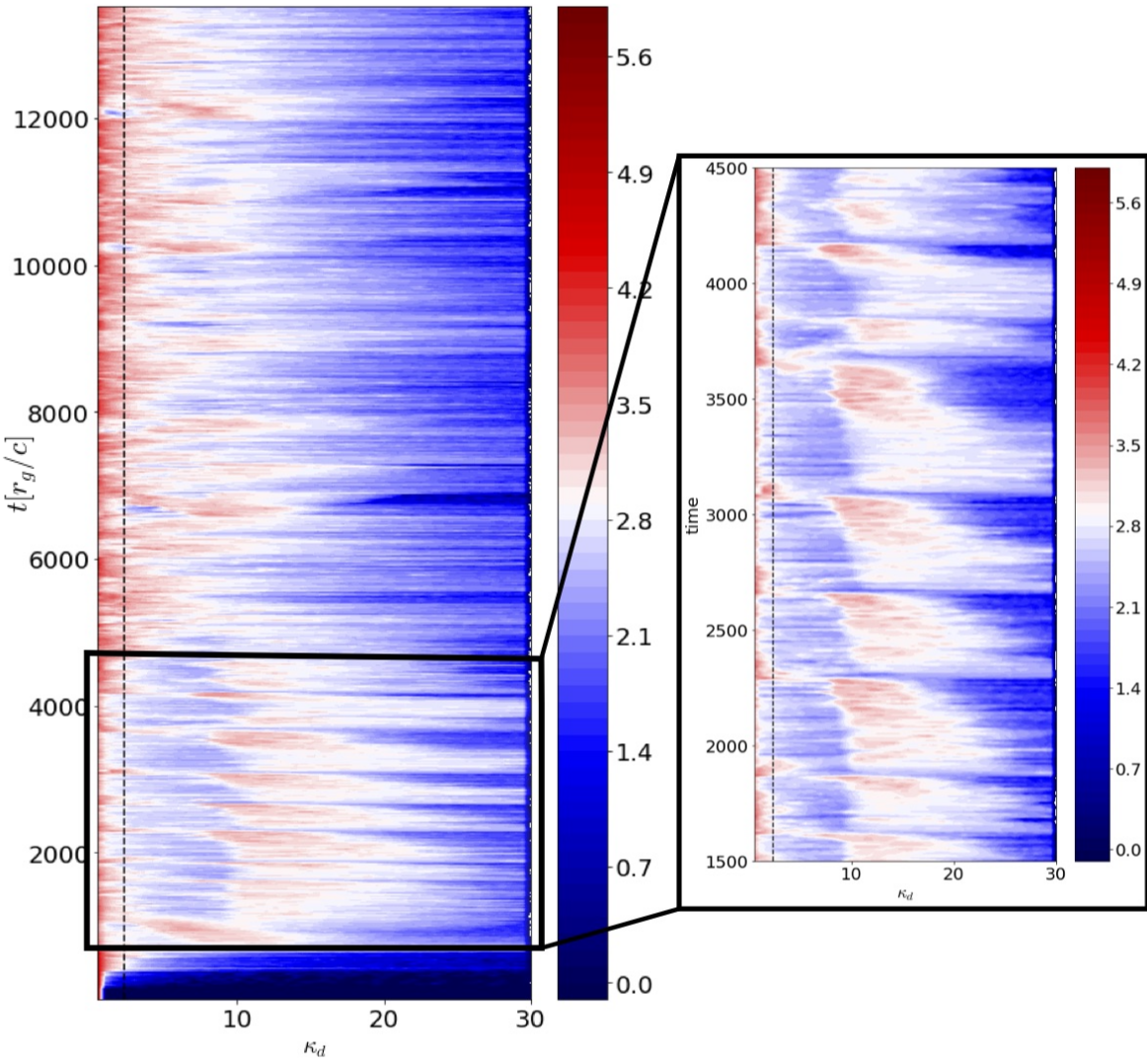}
    \caption{Time-resolved histogram of   characteristic length scales of $B_{\rm p}$ in the disc region  in model {\tt 2D1}. Black dashed line shows $R_{\rm ISCO}$. The colors show the logarithmic of the number of cells with scale $\kappa_d$ inside of a sphere with a radius $r\leq R_{\rm ISCO}$.}
    \label{fig:figure4}
\end{figure*}

The main mechanism of energy extraction from a rotating BH proposed by \citet[]{BZ77} requires the existence of a large-scale magnetic field threading the horizon. 
In order to analyze the efficiency of the BZ process, we estimate the length scales of the magnetic fields.
For this we use a characteristic length scale
\begin{equation}
    \kappa=\displaystyle \frac{|{\mathbf B_{\rm p}}|}{[\nabla \times {\mathbf B}]_{\phi}},
\end{equation}
which gives a measure to the radius of the region containing the currents generating the poloidal magnetic field $\mathbf{B}_{\rm p}$. 
This quantity has different typical values in the MAD and SANE states. In Fig.~\ref{fig:figure4} we show  time-resolved histograms for $\kappa$, where the black dashed line marks the scale of $R_{\rm ISCO}$. 
The histogram was calculated in the disc region ($45^\circ<\theta<135^\circ$) and it marked with $\kappa_d$.
We calculate $\kappa_d$ in all cells between the horizon and $R_{\rm ISCO}$ and bin the result into 60 evenly distributed bins.
The $x$ axis shows the value of $\kappa_d$ and the colors designate the number of cells heaving this value in $\log$ scale.
During the MAD state ($1000\tg<t<4500\tg$) the distribution is bimodal with a second pick at small scales ($\kappa_d<R_{\rm ISCO}$), which arise from the formation of plasmoids due to reconnection on the equatorial plane. 
In SANE state ($t>4500\tg$), the disc histogram is dominated by small-scale magnetic fields accreted onto the BH with the accretion flow. 

\subsubsection{Efficiency}
The efficiency can be defined as the temporal power output relative to the average power supplied to the BH by the disc. We use two definitions of efficiency associated with different forms of energy.
The first measures the total power output from the system relative to the rest-mass energy flux that crosses the horizon, and is defined as:
\begin{equation}
        \eta\uu{C}=\displaystyle\frac{F\uu{M}-F\uu{E}}{<F\uu{M}>}\times 100\%,
\end{equation}
    where $F\uu{M}=\dot Mc^2$ is the rest-mass energy flow through the horizon, $<F\uu{M}>$ is the average of $F\uu{M}$ over some time interval $\Delta t_{\rm av}$, shown in  Tables~\ref{tab:2D_simulations}, and \ref{tab:3D_simulations} and $F\uu{E}=\iint T^r_t {\rm d}A\uu{\Omega}$ is the momentum flow on the horizon with $T^r_t$ the $rt$ component of the stress-energy tensor $T^{\mu}_{\nu}$. This definition matches the one at \citet{Tchekhovskoy2012}.
    The second definition of efficiency measures the electromagnetic (EM) power output from the BH horizon relative to the maximal BZ output from the magnetic flux stored in a loop
   \begin{equation}
       \eta\uu{BZ}=\displaystyle\frac{F\uu{EM}}{F\uu{BZ_0}}\times 100\%,
   \end{equation}
   where $F\uu{BZ_0}=\displaystyle \frac{1}{24\pi^2 c}\Omega^2\uu{BH}\Psi_0^2$  \citep{Tchekhovskoy11}, $\Psi_0$ is the initial magnetic flux in the loop, $\Omega\uu{BH}=\displaystyle\frac{ac}{2r\uu{H}}$ is the BH angular frequency,
   and $F\uu{EM}=\iint [T\uu{EM}]^r_t {\rm d}A\uu{\Omega}$
   is the total EM power coming out of the BH horizon, with $[T\uu{EM}]^r_t=b^2u^ru_t-b^rb_t$ denoting the electromagnetic part of the $rt$ component of the stress-energy tensor.
   
\begin{figure}
    \centering
    \includegraphics[width=0.5\textwidth]{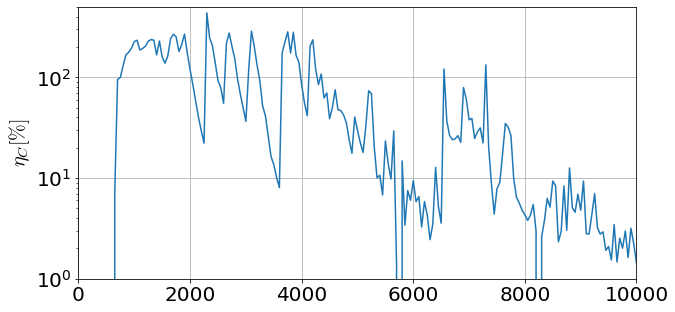}
     \includegraphics[width=0.5\textwidth]{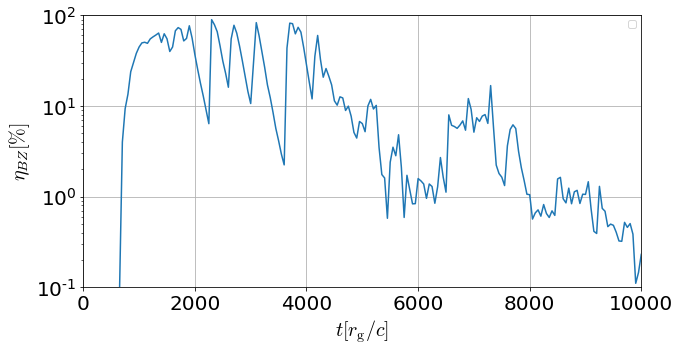}
    \caption{The evolution of the energy outflow efficiency (upper panel) and the jet efficiency (lower panel)
    in model {\tt 2D1}.}
    \label{fig:figure5}
\end{figure}

The efficiency $\eta\uu{C}$ measures the relative amount of accretion power that is lost to infinity from the entire system (BH+disc), while $\eta\uu{BZ}$ measures the fraction of the stationary BZ power (associated with initial magnetic flux) 
that emerges from the BH along open field-lines via the BZ effect, most of it is channeled into the jet.
 We show both efficiencies in Fig.~\ref{fig:figure5}. 
 The jet efficiency correlates with the flux through the horizon as expected. The jet power is maximal during the MAD state. It drops after $t\sim 4500~\tg$ due to two effects: (1) the magnetic field inside $R\uu{ISCO}$ becomes dominated by small size loops accreted from the disc, rendering the BZ process inefficient. (2) The mass accretion rate (and the horizon magnetic flux) drops considerably due to the limited amount of matter in the torus. In the MAD regime both $\eta\uu{C}$ and $\eta\uu{BZ}$ behave similarly, but in the beginning of the SANE regime ($t=4500-7000~\tg$) the outflow efficiency $\eta_C$ remains high because of strong disc wind. 
 
\subsubsection{The influence of global parameters}

The simulation results are influenced by the global parameters of the model, among them are the initial magnetization and magnetic field configuration, the equation of state (EOS) and the disc and Kerr BH parameters. We run variants of model {\tt 2D1} where we changed some of its parameters (models {\tt 2D1a}, {\tt 2D1b}, {\tt 2D1c} and {\tt 2D1d} in Table \ref{tab:2D_simulations}, links to movies show the evolution of each model are available in tables as well) and describe the main effects in each variant. 

\begin{itemize}
    \item Equation of state (model {\tt 2D1a}): Stiffer EOS with $\gamma=5/3$ tends to increase the disc thickness because matter becomes less compressible and reduces the magnetization of the accreted material. The second effect may be understood by considering a volume deformation over a characteristic length scale $L$ in the ideal MHD limit. Since density changes as $\rho\propto L^{-3}$, gas 
    pressure changes as $p_{g}\propto L^{-3\gamma}$, while magnetic pressure follows $B^2\propto L^{-4}$ in a case of an isotropic chaotic magnetic field. From this we deduce that $\beta$ scales as $\beta\propto L^{4-3\gamma}$. In the case where $\gamma=4/3$ the ratio of gas to magnetic pressure is constant, while for $\gamma=5/3$ the inflowing matter becomes more gas pressure dominated, as $\beta\propto L^{-1}$.
    We find that these effects lead to two outcomes: i) the characteristic duration of the relaxation cycle in a MAD state increases and ii) the system remains in a MAD state during the entire simulation and does not switch to a SANE mode as in the fiducial run with $\gamma=4/3$.  
    
    \item Initial disc magnetisation: Larger $\beta_{\rm min}$ (model {\tt 2D1b}) increases the magnetic amplification time and decreases the duration of the MAD regime.  In the case of extremely large $\beta_{\rm min}\sim 1000-10000$ there is no MAD regime at all. The reason for that is likely the small discs we use in our simulations, which run out of matter by the time the horizon flux becomes sufficiently high. 
    \item In the case of a counter-rotating black hole (model {\tt 2D1d}), the magnetic flux is accumulated faster at the beginning of the simulation as the last stable orbit radius in this case is larger than the co-rotation radius, thus the initial accretion rate is higher. The duration of the MAD stages is similar to the duration in the co-rotating case.
    \item In our simulations, the disc size was comparable to the size of the simulation box. To verify that the results are not affected by the box size we run a simulation with a much larger box (model {\tt 2D1c}).  We find that increasing the outer boundary radius
    has little effect on the results, at least over times $t<10000 \tg$.
    
\end{itemize}

\subsection{2D multiple loops models}

All accreting BH systems exhibit variable emission over a range of timescales. Some scales are related to the characteristic size of the system or to the dynamics of the accreted plasma, and some are related to the typical scale of the magnetic field in the disc. 
In real discs magnetic fields are chaotic and are amplified through turbulence and MRI. Their spatial scales are likely to be on the order of the disc thickness, and there is no reason to assume that in the majority of cases a single loop grows and dominates the accretion process (though see \citealt{Liska2020}).
To test the effect of multiple magnetic loops on the accretion flow and on the jet launching we tested several configurations of magnetic field loops, which we classify into two types.  The first one, termed {\it dipole} configuration consists of  alternating-polarity loops centered on the equatorial plane. In the second, termed {\it quadruple},  the magnetic loops are located above and below the equatorial plane and have opposite polarity (see Eq.~\ref{eq:2}). 
The different configuration types are shown in Fig.~(\ref{fig:figure1}).
We compared the efficiency of energy extraction in the fiducial model {\tt 2D1} to a dipole configuration with 3 loops (model {\tt 2D3}) and a quadruple configuration with two loops (model {\tt 2D2}). 

\begin{figure}
    \centering
    \includegraphics[width=0.5\textwidth]{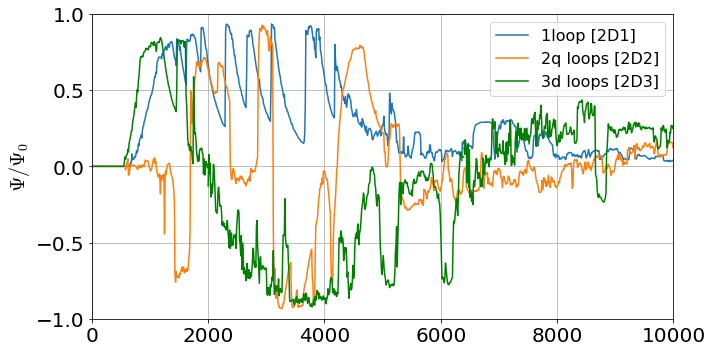}
    \includegraphics[width=0.5\textwidth]{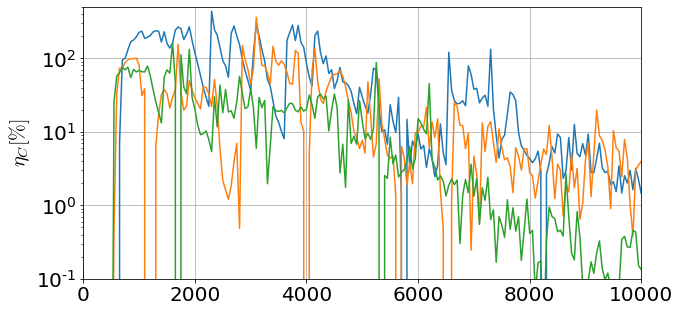}
    \includegraphics[width=0.5\textwidth]{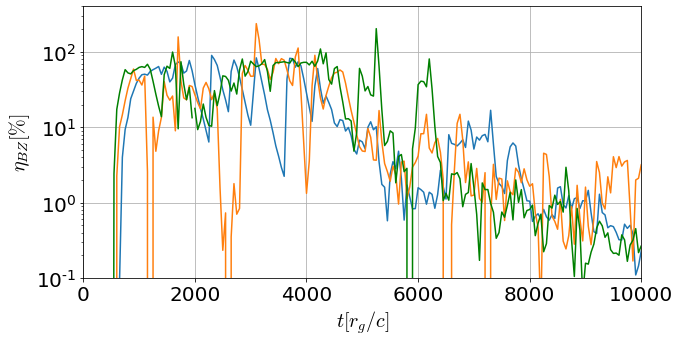}
    \caption{The evolution of the dimensionless magnetic flux through the horizon (upper panel), the energy outflow efficiency (middle panel) and the jet efficiency (lower panel) in models {\tt 2D1} (blue), {\tt 2D2} (orange) and  {\tt 2D3} (green). The values of $\Psi$ and $\eta_{\rm BZ}$ are normalized according to the initial flux and total magnetic energy in the loop that is currently being accreted.}
    \label{fig:figure6}
\end{figure}

The upper panel in Fig. \ref{fig:figure6} shows the time evolution of the flux on the horizon in the three configurations. The flux at each time is normaized with respect to the initial flux of the loop currently being accreted. 
Accretion in all models begins at roughly the same time. In models {\tt 2D1} and {\tt 2D3} the accretion is accompanied by an increase of magnetic flux on the horizon. 
The flux rises faster in model {\tt 2D3} since it is initially confined to a loop of a smaller size and thus accumulates faster. The system then evolves to a MAD state that persists throughout the accretion of the first loop with little influence by the second loop, which awaits to be accreted. When the 
mass associated with the first loop accretes, the pressure applied by the second loop compresses the field lines of the first loop causing them to reconnect and to release the energy stored in them. The flux on the horizon drops to zero and accretion of the second loop begins. As a result the horizon magnetic flux is completely replaced by the flux of the second loop, which has an opposite sign.
As can be seen in Fig. \ref{fig:figure}, most of the flux of the initial loop reaches the horizon, indicating a small dissipation rate from the onset of accretion. The accumulation of flux on the horizon leads to a second MAD state and to the launching of a second double sided jet with an opposite polarity. In our simulation the initial flux of the second loop is two times smaller than the flux of the first loop and  the jet power is smaller by a factor of four. The accretion of the third loop again leads to the reconnection of the second loop magnetic field, which makes way for the coming field with opposite polarity. However, here the advected flux is not high enough to start a MAD state and accretion proceeds in SANE mode.

\begin{figure}
    \centering
    \includegraphics[width=0.5\textwidth]{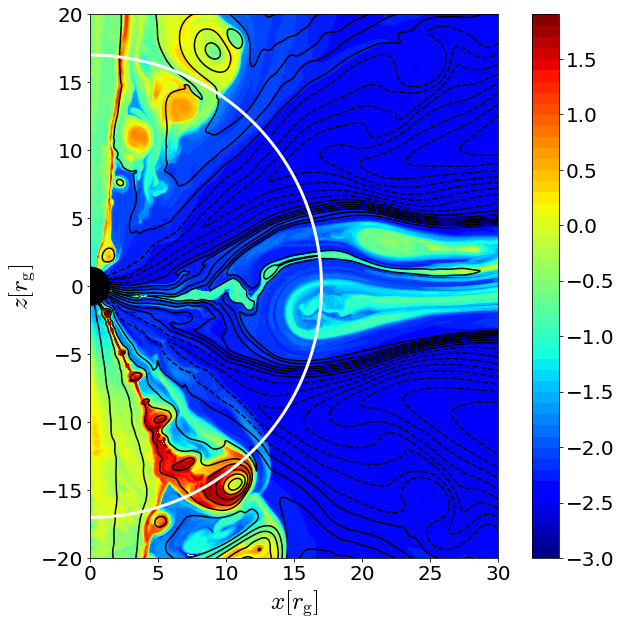}
    \includegraphics[width=0.5\textwidth]{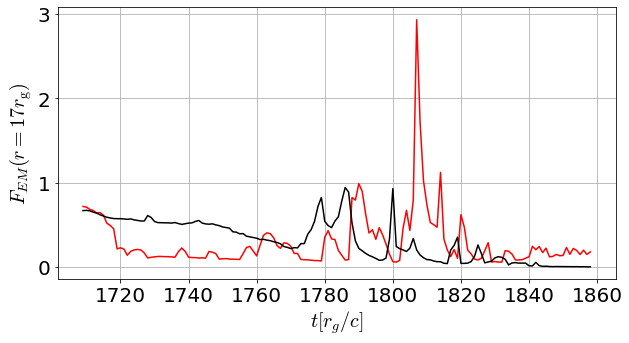}
    \caption{The formation of plasmoids in current sheets during the time when the second loop begins to be accreted in model {\tt 2D3H}. Top panel shows a snapshot of the entropy ($\log (p/\rho^{\gamma})$) at time $t=1811\tg$. Magnetic fields with opposite polarity are shown in solid and dashed lines. Current sheets are formed on the equatorial plane and along the boundary between the jet and the disc wind. Plasmoids are formed in the current sheets in region where increase dissipation occurs and are seen as high-entropy regions in the plot.  
    Bottom panel shows the time evolution of the electromagnetic energy flux in the jet (black curve) and outside of the jet (red curve) measured at radius $r=17r_{\rm g}$ (shown as a white circle). The peak in the red curve at time $t=1811 \tg$ is caused by the passage of a large plasmoid through the measured surface.}
    \label{fig:figure7}
\end{figure}

During the reconnection episodes that accompany the change in polarity we witness enhanced formation of small scale plasmoids. We see the plasmoids in the high-resolution simulation (model {\tt 2D3H}). These plasmoids are formed on the interface between the jet and the disc wind and on the equator, where current sheets are formed, as seen in Fig.~\ref{fig:figure7}.
The plasmoids carry energy comparable to or grater than the jet energy at that time (Fig.~\ref{fig:figure7}).  In Fig~\ref{fig:figure11} (left panel) we show the time resolved histogram of $\kappa\uu{j}$ in the jet  regions ($\theta<45^\circ$ and $\theta>135^\circ$) for model {\tt 2D3H}. During the MAD episodes the distribution of $\kappa\uu{j}$ is dominated by scales larger than $R\uu{ISCO}$. The enhanced states of plasmoid formation take place between MAD states and are characterized by a peak in the $\kappa\uu{j}$ distribution at scales smaller than $R\uu{ISCO}$.  We use this characteristic  distribution in the 3D case (Fig.~\ref{fig:figure11}) to identify the plasmoid formation periods similar to the 2D case. These episodes are clearly seen in the movies of \href{https://www.youtube.com/watch?v=Bk-OgPn-lQc}{the dipole} and the \href{https://www.youtube.com/watch?v=eQcvWEnM2G0}{quadruple}  configurations before or during switching of polarity at the BH horizon. 

The quadruple configuration  (model {\tt 2D2}) is initially anti-symmetric with respect to the equatorial plane.  Accretions  
starts at the same time as in the dipole configuration, and both loops reach the BH horizon simultaneously. The accretion of each loop takes place on a single hemisphere\footnote{As oppose to be accreted on the entire BH horizon as in the dipole case}, thus the total flux on a hemisphere remains zero and does not reflect the power output from the BH, as can be seen in Fig.~\ref{fig:figure6} and in \href{https://www.youtube.com/watch?v=eQcvWEnM2G0}{the movie}.
A double-sided jet is formed with efficiency comparable to the 3 dipole loops case (model {\tt 3D3}).
At $t\sim 1500 \tg$ the symmetry breaks as the southern loop pushes aside the northern loop and begins to accrete over the entire BH horizon. At this stage a regular MAD state is ewached, similar to the dipole case. The accretion switches between the two loops, where at each time a different loop blocks the accretion flow of the other, goes into a regular MAD state and launches a mini jet with an alternating magnetic polarity.  
The total duration of the MAD phases and the energy released are determined by the initial loops size and by their flux, which are comparable to the single loop case.
In this configuration we also witness plasmoid chains growing in current sheets that form during periods of polarity switch on the BH horizon. However the morphology in this case is more complex than in the case of the dipole configurations as can be seen in \href{https://www.youtube.com/watch?v=eQcvWEnM2G0}{this movie}.


\subsection{3D multiple loops models}

\begin{figure}
    \centering
    \includegraphics[width=0.5\textwidth]{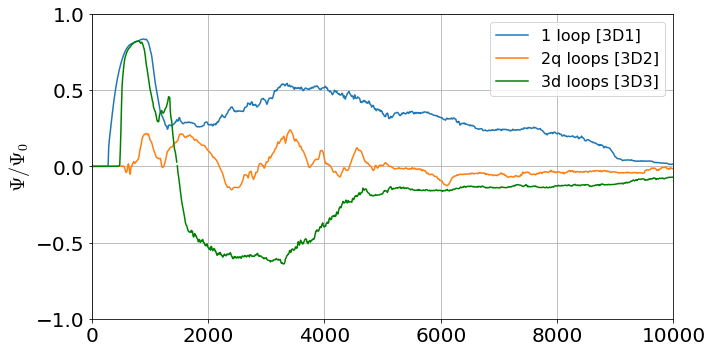}
    \includegraphics[width=0.5\textwidth]{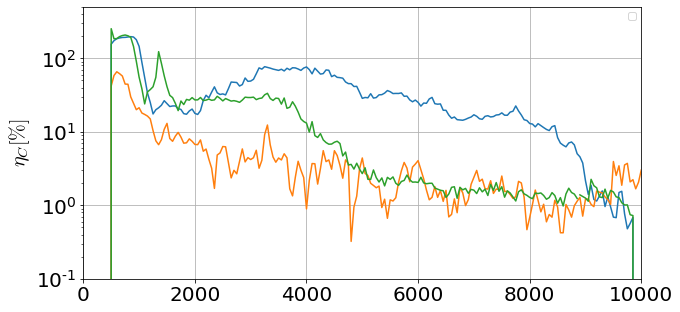}
    \includegraphics[width=0.5\textwidth]{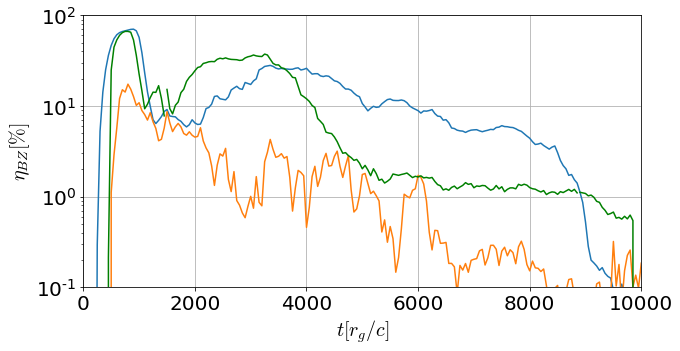}
    \caption{The evolution of the dimensionless magnetic flux through the horizon (upper panel), the energy outflow efficiency (middle panel) and the jet efficiency (lower panel) in models {\tt 3D1} (blue), {\tt 3D2} (orange) and {\tt 3D3} (green). The values of $\Psi$ and $\eta_{\rm BZ}$ are normalized according to the initial flux and total magnetic energy in the loop that is currently being accreted.}
    \label{fig:figure8}
\end{figure}

Most phenomena associated with magnetic fields, such as the dynamo process or MRI are essentially three-dimensional. 
Therefore, it is important to model the evolution of the system in 3D in order to asses the relevancy of our 2D results. 
We run simulations with the same initial conditions and magnetic field configurations as in the 2D models. The models are listed in table~\ref{tab:3D_simulations}. 
The computational box has $r_{out}=10^5r_g$ and 
we used a fiducial resolution of $512\times 256 \times 64$ cells.
Fig. ~{\ref{fig:figure8}} shows the evolution of the magnetic flux through the horizon and of the two efficiency types we defined above, in the same configurations tested in section 2 (see Fig. \ref{fig:figure6} for a comparison). Quite generally all three parameters evolve more smoothly in the 3D runs than in 2D. 

The accretion in all 3D models begins somewhat earlier than in the 2D case, likely due to a combination of a higher numerical resistivity because of lower resolution, and a more efficient magnetic field amplification. 

In the case of a single loop and three dipole loops, flux buildup leads to MAD states with comparable outflow ($\eta\uu{C}$) and jet ($\eta\uu{BZ}$) efficiencies to the 2D cases of about $100\%$. The dynamics during the MAD state, however is different. 
Unlike in the 2D case (Fig \ref{fig:figure3}), here we do not see the intermittent behaviour induced by the reconnection on the equatorial plane. Instead the accretion process is smoother and shows only small amplitude variations. The reason for that is a combination of two factors. First, the reconnection on the equatorial plane is not uniform, and mass continues to be accreted from regions with lower reconnection rate, producing a spiral pattern of high density and low density accretion streams. 
Second, the low resolution in the 3D case leads to high effective resistivity, allowing matter to diffuse across detached field-lines in regions with high reconnection rate, thus accretion is not completely halted even in those regions.  Further high resolution 3D studies are needed to resolve the complex dynamics and will be performed in a follow-up work.

\begin{figure}
    \centering
    \includegraphics[width=0.5\textwidth]{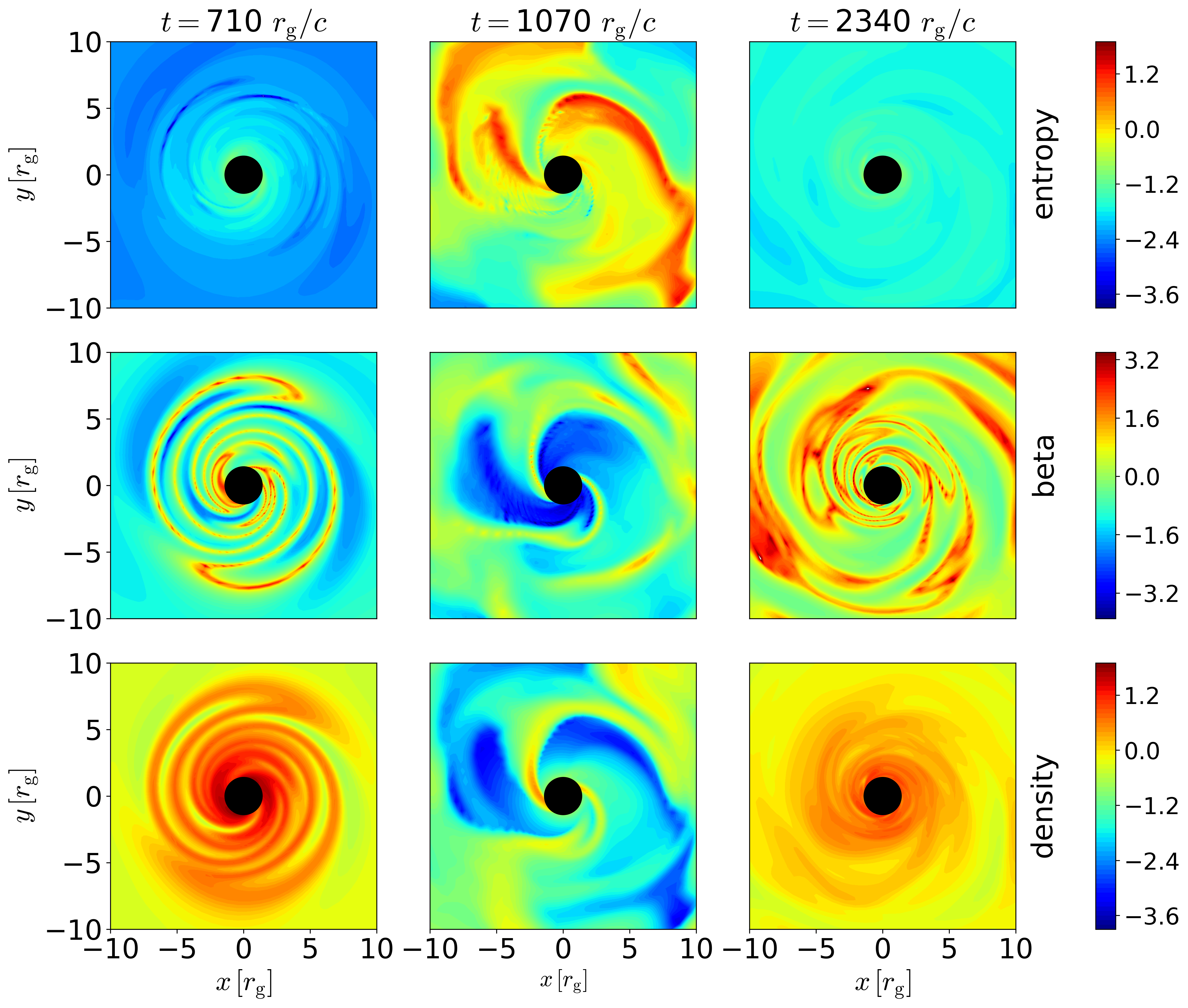}
    \caption{From top to bottom, entropy, $\beta$, and density distributions on the equatorial plane  before, during and after a major reconnection episode in model {\tt 3D3}. The peak of the reconnection occurs at time $t=1070\,\tg$ (middle column). During this time  the entropy is maximized  while the density and $\beta$ are at their minimum. Since the reconnection is not uniform on the plane mass continuous to be accrete through the regions where reconnection rate is low. This phase of low accretion lasts a few hundreds dynamical times. 
    }
    \label{fig:figure9}
\end{figure}

In Fig.~\ref{fig:figure9} we show an example of a major reconnection event seen in model {\tt 3D3} at time $t=1090~\tg$, which removes a sizable fraction of the flux on the horizon leading to a drop in the accretion rate and in the values of both efficiencies. 
The accretion is resumed shortly after and with it the horizon flux and efficiencies spike up, similar to the intermittent pattern seen in 2D. Such episodes seem to be rare and occur randomly in the 3D case unlike the regular reconnection behaviour seen in the 2D case. Note that a similar drop in the flux and accretion rate is observed in the single loop case (model {\tt 3D1}). However here it remains low for $\sim2000~\tg$. The reason for that is likely to be the high effective resistivity which allows matter to slip across fieldlines, preventing accumulation of mass and the consequent exertion of sufficient ram pressure to close the gap. 
High resolution 3D simulations of a single loop by Ripperda et. al. 2021 show similar overall accretion and reconnection patterns as in our model {\tt 3D3} but with a more regulated gap dynamics. 
In the three-loops case the pressure from the second loop that approaches the horizon helps closing the gap faster. This pressure eventually forces reconnection of the first loop, leading to a complete switch in the flux direction on the BH horizon at $t=1700~\tg$, as in the 2D case. The efficiencies during the accretion phase of the second loop ($1700\tg \lesssim t\lesssim5000\tg$) in model {\tt 3D3} are somewhat lower than in the 2D case, likely due to the lack of the regular MAD behaviour. Ohmic dissipation, likely enhanced due to the low resolution, leads to a loss of flux in the third loop thus the system enters a low horizon flux SANE state at later times when the accretion of the second loop ends. 

The behaviour of the 3D quadruple configuration (model {\tt 3D2}) shows some similarities to the 2D case, however the energy extracted from the system is much lower.
In the 2D case the two quadruple loops, initially located on both sides of the equatorial plane, push each other sequentially, so that at each time a single loop threads the entire horizon. The loop reaches a MAD state and launches a two-sided jet with a matching magnetic field polarity, before it is replaced by the second loop with opposite polarity. In the 3D case, the contact surface between the loops oscillates about the equatorial plane, but generally remaining at a low latitude. Thus unlike in the 2D simulations, accretion on each hemisphere proceeds on a separate loop. In such a case the accreted material fills a larger fraction of the horizon leading to a SANE like accretion behaviour with turbulent structure of small-scale magnetic fields and multiple current sheets. 
This dynamics is captured in the scale distribution of the magnetic field shown in Fig.~\ref{fig:figure11}a, as an over-abundance of cells with small scale magnetic fields at $r<R\uu{ISCO}$. 
As a result both the total magnetic flux on the horizon as well as the two types of efficiencies are much lower than in the 2D case. 
In addition, 
the magnetic field frozen into the matter is dragged along with the turbulent inflow through the horizon thereby decreasing the overall output of electromagnetic energy. 
We can see this process as dips in efficiency in the third panel of Fig.~{\ref{fig:figure8}}.

\begin{figure}
    \centering
    \includegraphics[width=0.5\textwidth]{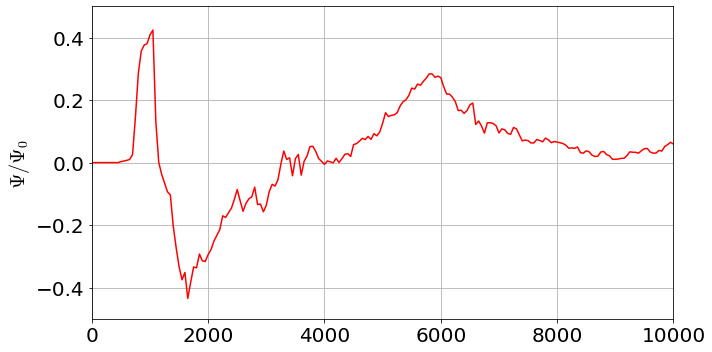}
    \includegraphics[width=0.5\textwidth]{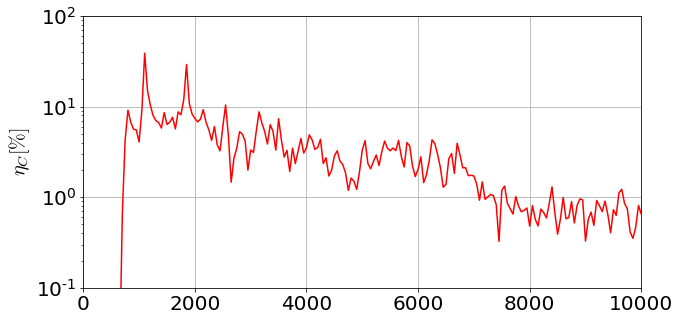}
    \includegraphics[width=0.5\textwidth]{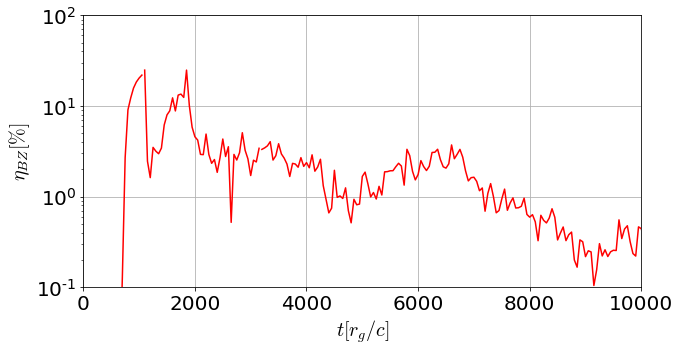}
    \caption{ The evolution of the dimensionless magnetic flux through the horizon (upper panel), the energy outflow efficiency (middle panel) and the jet efficiency (lower panel) in model {\tt 3DM}. The values of $\Psi$ and $\eta_{\rm BZ}$ are normalized according to the initial flux and total magnetic energy in the loop that is currently being accreted.}
    \label{fig:figure10}
\end{figure}

\begin{figure}
    \centering
       \includegraphics[width=0.5\textwidth]{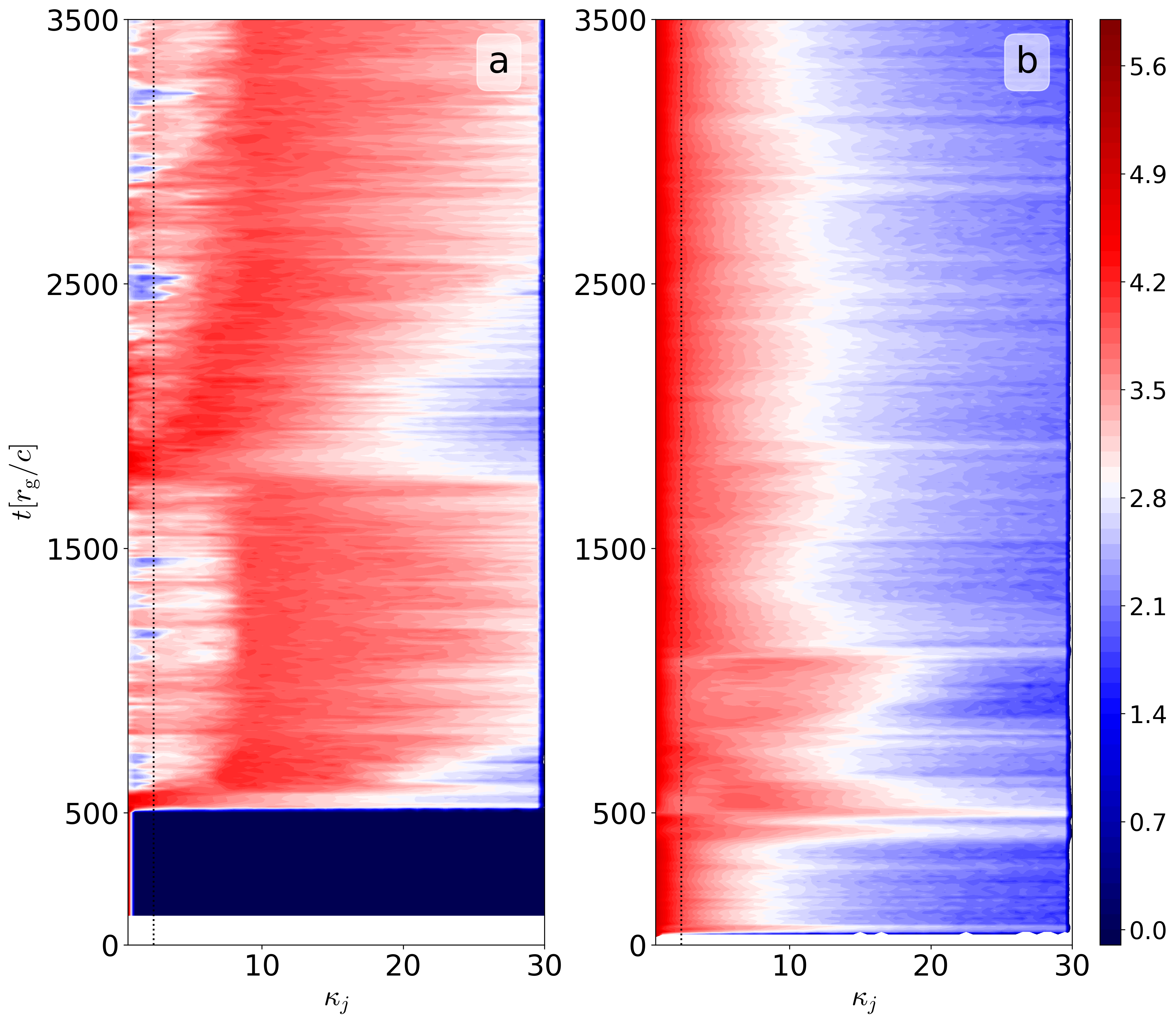}
    \caption{Time-resolved histogram of characteristic length scales of $B_{\rm p}$ in the jet region  in models {\tt 2D3H} (panel a) and {\tt 3DM} (panel b). Black dashed line shows $R_{\rm ISCO}$. The color show the logarithmic of the number of cells with scale $\kappa_j$ inside of a sphere with a radius $r\leq R_{\rm ISCO}$.}
    \label{fig:figure11}
\end{figure}

\begin{figure*}
    \centering
     \includegraphics[width=\textwidth]{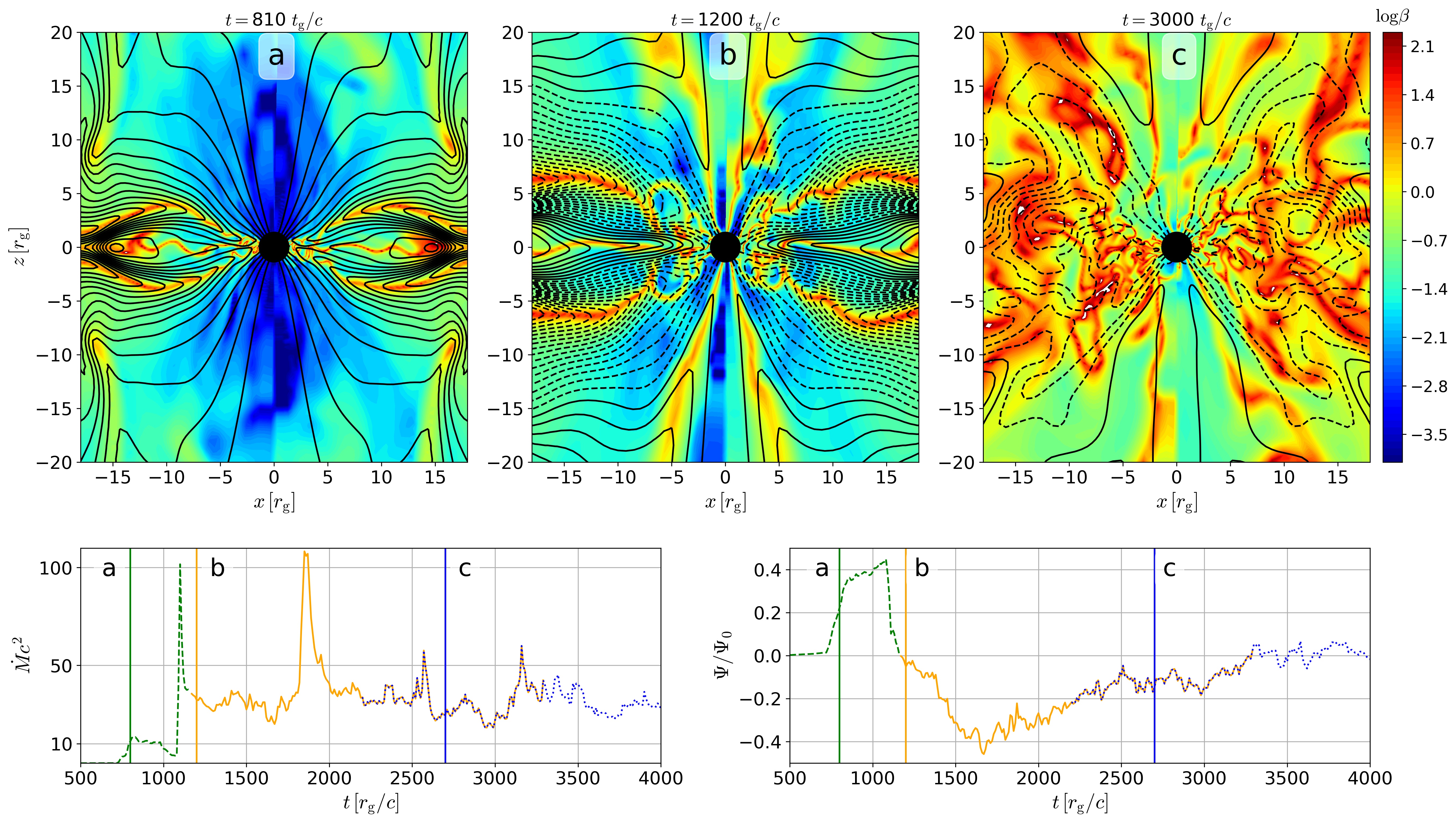}
    \caption{
    The accretion dynamics in model {\tt 3DM}. The three upper panels show snapshots of $\log \beta$. Magnetic field lines are shown in black lines. The lower panels show the mass accretion rate at the horizon (left) and normalized magnetic flux through horizon (right). Vertical lines indicate the times of the snapshots.
    }
    \label{fig:figure12}
\end{figure*}

In order to test the effect of multiple polarity switches on the BH horizon, 
we ran a simulation with a large number of loops having identical initial magnetic fluxes (model {\tt 3DM}). The evolution of the system is show in  Fig.~\ref{fig:figure10}). It can be seen that the efficiencies and the magnetic flux during the accretion of the first two loops in this case are smaller than in models {\tt 3D1 and 3D3}. The reason for that is likely because the loop size is too small and the system does not time to reach a full MAD state before the next loop with an opposite polarity comes in. The late evolution of the system is similar to the other dipole models since the system enters a turbulent SANE state.  The evolution of the system during the accretion of the first two loops is shown in details in Fig.~\ref{fig:figure12}. We show the mass accretion rate (left bottom). The horizon flux  (right bottom) the accretion of each loop is highlighted by a different color and line style.  At the top panels we show snapshots from three distinct times during the accretion of the first and second loops. These times are marked with solid vertical lines at the bottom panels.
Mass starts to accrete on the BH at $t=750~\tg$ following a buildup of magnetic field in the disc. The flux buildup on the horizon leads to a short MAD state ($800\lesssim t\lesssim1100~\tg$) accompanied by the launching of a double-sided Poynting flux dominated jet via the BZ process (panel a). Inside the last stable orbit magnetic stresses dominate over the accretion flow ram pressure and the accretion rate decreases.
The MAD state is terminated by a jump in the mass accretion rate driven by the pressure of the second loop as it approaches $R\uu{ISCO}$. It leads to large-scale reconnection of magnetic flux on the horizon followed by a switch in the field polarity at $t=1200~\tg$. The switching of  polarity is accompanied by the formation of current sheets both in the accretion torus and on the jet boundary (panel b). In these current sheets plasmoids are created due to magnetic reconnection. The process is clearly visible in 2D simulations, but due to poor resolution, we cannot see the plasmoids themselves in 3D simulations. We are able to infer their presence from the distribution of magnetic curvature radii shown in Fig.~\ref{fig:figure11}. During the polarity switching, small-scale magnetic fields prevail inside the last stable orbit. Based on our 2D results we associate that with plasmoid formation. The accretion of the second loop leads to a second MAD state and to the launching of a Poynting flux dominated jet with a matching polarity. This state
ends again by the crushing of matter on the the BH at $t\simeq1800~\tg$ due to pressure by the third loop as it approaches the ISCO radius. In this case, however, the flux switched only close to the poles, while at low latitudes mass associated with the second loop continues to be accreted (panel c). A complete flux switch occurs only at $t\simeq3300~\tg$. However by that time turbulence and Ohmic dissipation have destroyed the large scale structure of the loop, leading to a SANE state, which continues throughout the rest of the simulation.
All together during the accretion of the first three loops the three highly magnetized mini jets were launched carrying counter-oriented magnetic fields, while the rest of the loops broke apart in the disc and were unable to generate jets.

\section{Conclusions}
In this paper we explore activation channels of a Kerr black hole by accretion of small scale magnetic fields.
We focus on initial magnetic field configurations consisting of dipole and quadruple loops with alternating polarity.
Such configurations are expected to give rise to a rapid dissipation in current sheets that form on horizon scales, as well
as to quasi-striped jets \citep{Parfrey15,mahlmann20}. 
To that extent, we performed 2D and 3D GRMHD simulations and examined the accretion dynamics, dissipation pattern and jet launching efficiencies in 4 different cases: a single loop, two quadrupolar loops, three dipole loops and multiple dipole loops. Our main findings are:

\begin{itemize}
\item We identified cyclic behaviour of the accretion dynamics in MAD states that alternate between episodes of 
high accretion rate to episodes of quenched accretion during which the magnetosphere becomes nearly entirely 
force-free out to radii $\gtrsim 10 r_g$, with a thin equatorial current sheet. 
This cyclic behaviour suggests that the observed characteristics (notably, peak luminosity, spectrum, shadow image, etc.) of the emission from the
innermost magnetospheric regions (within $\sim 10 r_g$) should vary over timescales of tens to hundreds $\tg$ (weeks to months in M87). In particular, 
we contend that the emissivity profile anticipated in the inner magnetosphere during the states of quenched accretion 
may be inconsistent with the BH shadow image reported by the EHT collaboration \citep{EHT2019p1,EHT2021}. Moreover, we anticipate enhanced high energy emission from the current sheet during these states, although quantifying its properties would require a separate analysis. 
It is worth noting that GRMHD simulations invoke artificial plasma injection (density floor) in low density regions, and the question remains as
to how plasma is supplied to the jet base.  Continuous plasma supply may be 
contributed by spark gaps \citep{Levinson11,levinson18,kisaka20,crinquand20}. However, this requires the soft-photon intensity to exceed a certain level,
and it is unclear at present whether emission during the quenched accretion states can accommodate that.  Alternatively, it could be that 
 jet formation is suppressed during these states.

\item We find that advection of dipole loops with alternating polarity can lead to formation of a quasi-striped jet with substantial mean power (around $10-20$\% of the jet power created in the single loop model), provided the characteristic loop size is not too small.  This is qualitatively consistent with results of GRFFE (force-free) simulations \citep{Parfrey15,mahlmann20}.  However, it requires the loops to form or be maintained at a radius $r \lesssim 100 r_g$. In our GRMHD simulations (like in most other similar works) matter is not injected into the simulation box through the outer boundary, 
as expected in reality. Therefore, we cannot have a persistent state over times longer than the advection time of 2 or at most 3 loops.  In addition, limited resolution and the lack of self-consistent cooling, that can affect the disc structure and magnetization in yet unknown ways, are likely to have an effect on the launched jet. Nonetheless, our results indicate that striped jets with significant power can potentially form, enabling a natural dissipation channel for magnetically dominated jets, particularly in AGNs, where conditions for generation of instabilities that can destroy the magnetic field seem less likely.

\item Our simulations confirm that the advection of the field onto the BH leads to enhanced dissipation in current sheets and to the formation of plasmoids in various locations in the vicinity of the BH: i) near the equatorial plane, during episodial reconnection events characterizing MAD states, ii) along the jet boundary during the time when the accreted loop reconnects upon itself  making way to the next loop advected onto the BH.
The plasmoids in both cases were directly observed in the 2D simulations, while in the 3D simulations their existence was inferred from measurements of the curvature radii of fieldlines inside the ISCO radius. 
These current sheets might be associated with the compact corona that produces the hard X-ray emission seen in many 
Galactic and extra-galactic accreting BH systems (as in "lamppost" models, \citealt{Yuan19,mahlmann20,sironi20}).  
These dynamical structures should not, necessarily, be associated with jet launching. They can form also under conditions not suitable for jet formation, as in our multiple loops model. 
\end{itemize}

We stress that our 3D simulations were not converged. Although they showed overall behaviour similar to the 2D simulations and to other high resolution 3D simulations that were run with a single loop configuration (Ripperda et. al. 2021), a proper convergence study needs to be made to validate the results. Such a study is beyond our current computational capabilities and will be preformed elsewhere.  

\section*{Acknowledgements}

We thank A. Philippov and B. Ripperda for useful discussion and for sharing their results with us. AL acknowledge support by the Israel Science Foundation grant 1114/17. OB and AC were supported by ISF grant 1657/18 and by ISF (Icore) grant 1829/12.

\bibliographystyle{mnras}
\bibliography{mybib} 
\bsp	
\label{lastpage}
\end{document}